\documentclass[a4paper,12pt]{article}
\pdfoutput=1
\usepackage{epsfig}
\usepackage{amssymb}
\usepackage{amsfonts}
\usepackage{amsmath}
\usepackage{euscript}
\usepackage{verbatim}
\usepackage{latexsym}
\usepackage{graphicx}
\usepackage{caption}
\usepackage{float}
\usepackage{subcaption}

\newif\ifdtup

\catcode`\@=11

\@addtoreset{equation}{section}

\def\@normalsize{\@setsize\normalsize{15pt}\xiipt\@xiipt
\abovedisplayskip 14pt plus3pt minus3pt%
\belowdisplayskip \abovedisplayskip
\abovedisplayshortskip \z@ plus3pt%
\belowdisplayshortskip 7pt plus3.5pt minus0pt}

\def\small{\@setsize\small{13.6pt}\xipt\@xipt
\abovedisplayskip 13pt plus3pt minus3pt%
\belowdisplayskip \abovedisplayskip
\abovedisplayshortskip \z@ plus3pt%
\belowdisplayshortskip 7pt plus3.5pt minus0pt
\def\@listi{\parsep 4.5pt plus 2pt minus 1pt
     \itemsep \parsep
     \topsep 9pt plus 3pt minus 3pt}}

\relax

\catcode`@=12

\catcode`\@=11

\def\section{\@startsection{section}{1}{\z@}{3.5ex plus 1ex minus
   .2ex}{2.3ex plus .2ex}{\large\bf}}

\def\SymBoxes#1#2#3#4{\newdimen\un@t \un@t#3%
\raisebox{#1}{\rule{#2\un@t}{#4}\hskip-#2\un@t
\@tempdimb\un@t \advance\@tempdimb by-#4\@tempcntb#2\relax%
\@whilenum{\@tempcntb>0}\do{
\rule{#4}{\un@t}\hskip\@tempdimb \advance\@tempcntb by\m@ne}%
\hskip-#2\un@t \rule[\un@t]{#2\un@t}{#4}%
\rule[\un@t]{#4}{#4}\hskip-#4
\rule{#4}{\un@t}}\hskip-#4}                

\begin{document}

\newcommand{\beq}{\begin{equation}}
\newcommand{\eeq}{\end{equation}}
\newcommand{\bea}{\begin{eqnarray}}
\newcommand{\eea}{\end{eqnarray}}
\newcommand{\beas}{\begin{eqnarray*}}
\newcommand{\eeas}{\end{eqnarray*}}
\newcommand{\defi}{\stackrel{\rm def}{=}}
\newcommand{\non}{\nonumber}
\newcommand{\bquo}{\begin{quote}}
\newcommand{\enqu}{\end{quote}}
\renewcommand{\(}{\begin{equation}}
\renewcommand{\)}{\end{equation}}
\def \eqn#1#2{\begin{equation}#2\label{#1}\end{equation}}

\def\e{\epsilon}
\def\IZ{{\mathbb Z}}
\def\IR{{\mathbb R}}
\def\IC{{\mathbb C}}
\def\IQ{{\mathbb Q}}
\def\de{\partial}
\def\Tr{ \hbox{\rm Tr}}
\def\H{ \hbox{\rm H}}
\def\HE{ \hbox{$\rm H^{even}$}}
\def\HO{ \hbox{$\rm H^{odd}$}}
\def\K{ \hbox{\rm K}}
\def\Im{ \hbox{\rm Im}}
\def\Ker{ \hbox{\rm Ker}}
\def\const{\hbox {\rm const.}}
\def\o{\over}
\def\im{\hbox{\rm Im}}
\def\re{\hbox{\rm Re}}
\def\bra{\langle}\def\ket{\rangle}
\def\Arg{\hbox {\rm Arg}}
\def\Re{\hbox {\rm Re}}
\def\Im{\hbox {\rm Im}}
\def\exo{\hbox {\rm exp}}
\def\diag{\hbox{\rm diag}}
\def\longvert{{\rule[-2mm]{0.1mm}{7mm}}\,}
\def\a{\alpha}
\def\dag{{}^{\dagger}}
\def\tq{{\widetilde q}}
\def\p{{}^{\prime}}
\def\W{W}
\def\N{{\cal N}}
\def\hsp{,\hspace{.7cm}}

\def\br{\nonumber}
\def\IZ{{\mathbb Z}}
\def\IR{{\mathbb R}}
\def\IC{{\mathbb C}}
\def\IQ{{\mathbb Q}}
\def\IP{{\mathbb P}}
\def \eqn#1#2{\begin{equation}#2\label{#1}\end{equation}}

\newcommand{\C}{\ensuremath{\mathbb C}}
\newcommand{\Z}{\ensuremath{\mathbb Z}}
\newcommand{\R}{\ensuremath{\mathbb R}}
\newcommand{\rp}{\ensuremath{\mathbb {RP}}}
\newcommand{\cp}{\ensuremath{\mathbb {CP}}}
\newcommand{\vac}{\ensuremath{|0\rangle}}
\newcommand{\vact}{\ensuremath{|00\rangle}                    }
\newcommand{\oc}{\ensuremath{\overline{c}}}
\newcommand{\psizero}{\psi_{0}}
\newcommand{\phizero}{\phi_{0}}
\newcommand{\hzero}{h_{0}}
\newcommand{\psiin}{\psi_{\rh}}
\newcommand{\phiin}{\phi_{\rh}}
\newcommand{\hin}{h_{\rh}}
\newcommand{\rh}{r_{h}}
\newcommand{\rb}{r_{b}}
\newcommand{\psibnd}{\psi_{0}^{b}}
\newcommand{\psibndp}{\psi_{1}^{b}}
\newcommand{\phibnd}{\phi_{0}^{b}}
\newcommand{\phibndp}{\phi_{1}^{b}}
\newcommand{\gbnd}{g_{0}^{b}}
\newcommand{\hbnd}{h_{0}^{b}}
\newcommand{\zh}{z_{h}}
\newcommand{\zb}{z_{b}}
\newcommand{\man}{\mathcal{M}}
\newcommand{\hbr}{\bar{h}}
\newcommand{\tbr}{\bar{t}}

\begin{titlepage}
\begin{flushright}
CHEP XXXXX
\end{flushright}
\bigskip
\def\thefootnote{\fnsymbol{footnote}}

\begin{center}
{\Large
{\bf A Hairy Box in Three Dimensions \\ \vspace{0.1in} 
}
}
\end{center}

\bigskip
\begin{center}
Chethan KRISHNAN$^a$\footnote{\texttt{chethan.krishnan@gmail.com}}, Rohit SHEKHAR$^a$\footnote{\texttt{rohitshekhar06@gmail.com}}, and  P. N. Bala SUBRAMANIAN$^b$\footnote{\texttt{pnbalasubramanian@gmail.com}}
\vspace{0.1in}

\end{center}

\renewcommand{\thefootnote}{\arabic{footnote}}

\begin{center}

$^a$ {Center for High Energy Physics,\\
Indian Institute of Science, Bangalore 560012, India}\\
\vspace{0.2in}
$^b$ {Institute of Mathematical Sciences, \\
	Homi Bhabha National Institute (HBNI),\\
	IV Cross Road, C.~I.~T.~Campus, \\
	Taramani, Chennai, 600113  Tamil Nadu, India }\\
\vspace{0.2in}
\end{center}

\noindent
\begin{center} {\bf Abstract} \end{center}
In this short note, we consider the phases of gravity coupled to a $U(1)$ gauge field and charged scalar in 2+1 dimensions without a cosmological constant, but with box boundary conditions. This is an extension of the results in arXiv:1609.01208, but unlike in higher dimensions, here the physics has sharp differences from the corresponding AdS problem. This is because Einstein-Maxwell black holes cease to exist when the cosmological constant goes to zero. We show that hairy black holes also do not exist in the flat 2+1 dimensional box under some assumptions, but hairy boson stars do. There is a second order phase transition from the empty box to the boson star phase at a charge density larger than some critical value. We find various new features in the phase diagram which were absent in 3+1 dimensions. Our explicit calculations assume radial symmetry, but we also note that the absence of black holes is more general. It is a trivial consequence of  a 2+1 dimensional version of Hawking's horizon topology argument from 3+1 dimensions, and relies on the Dominant Energy Condition, which is violated when (eg.) there is a negative cosmological constant.

\vspace{1.6 cm}
\vfill

\end{titlepage}

\setcounter{footnote}{0}

\section{Introduction}

Anti-de Sitter space is often compared to a box \cite{HP}, and recent work has shown that this analogy is quite robust: see \cite{CK, PCB} for two very different points of view. In particular, \cite{PCB, PCB-ads} showed that the phases of gravity coupled to a Maxwell field and a charged scalar in a flat space box are qualitatively very similar to that in AdS, if we are working in 3+1 (and presumably higher) dimensions. 

However, in 2+1 dimensions, we expect that this close parallel must receive some qualifications. The reason is that in asymptotically flat 2+1 dimensions, there are neither Schwarzschild black holes nor Reissner-Nordstrom black holes, 
whereas BTZ black holes \cite{BTZ} and their charged generalizations \cite{chargedBTZ} do exist in AdS$_{2+1}$. One might then ask the question: since a box can function as a proxy for a negative cosmological constant, can we find new solutions for flat space gravity in 2+1 dimensional gravity coupled to Maxwell and a charged scalar, by putting the system in a box?

At first sight, it might seem that there is nothing to be said regarding the 2+1 dimensional box. After all, Einstein's equations are local, and how can putting the system in a box far away, change anything that has to do with local physics? This non-argument misses a crucial point\footnote{Not to mention the fact that it has already been violated in 3+1 dimensions \cite{PCB} (see also various related previous and follow-up works in eg. \cite{Winstanley1, Winstanley2, Peng1}). But we feel it is worth a clarification, because this question has been raised on a few occasions.} which is that a box allows the possibility of non-trivial profiles for fields such that they have non-vanishing values at the boundary, thereby allowing the possibility of exotic new solutions. This is (often) impossible in asymptotically flat space, because it is natural for strong backreaction to result if the fields contribute too much energy-momentum in the asymptotic region. Indeed, in this paper we will see that it is possible to have new highly non-trivial solutions with lots of local structure in a 2+1 dimensional flat space box, for the Einstein-Maxwell-scalar system. The presence of the scalar is crucial, and without it there are no regular solutions. The solution space of these boson stars turns out to have interesting differences from higher dimensional cases. The study of their phase structure is the main goal of this paper.

But then one might even get bolder: since we know that allowing a box boundary provides a way to evade the no-hair theorems in 3+1 dimensions \cite{PCB}, is it conceivable that we could have a black hole supported entirely by hair in 2+1-d by putting it in a box? Despite the fact that we can put lots of localized matter and create boson stars, we will argue that having a black hole is impossible in a flat space 2+1-d box\footnote{Unless we allow potentials for the scalar, and effectively introduce a negative cosmological constant. We will discuss this caveat in more detail in section 5.} under fairly broad assumptions. It is possible to see the origins of this explicitly by looking at the perturbative structure of the equations of motion near horizon, in the spherically symmetric case. The technical reason is that one can trace the vanishing of the $g_{tt}$ piece in the metric (which can be viewed as necessary local requirement for the existence of a horizon) to a piece in the Einstein field equations: this piece is provided in AdS  by the cosmological constant, but in $(d+1)$-dimensional flat space it is provided by the curvature of the $(d-1)$-sphere. In 2+1 dimensions the relevant sphere is the circle which has no curvature, and therefore it cannot support such a solution. 

But the result is more general, and actually a trivial consequence of the 2+1 dimensional version of Hawking's 3+1 dimensional theorem that fixes the topology of the horizon \cite{Hawking, Oz}. With the same assumptions as in 3+1 dimensions, the theorem rules out horizons altogether in 2+1 dimensions! The fact that black holes are ruled out in 2+1 dimensions (under some assumptions) if the matter satisfies Dominant Energy Condition (DEC) has been noted before \cite{NoGo}.
These results are consistent with the view that a hairy black hole should really be thought of as a hairless black hole with hair added on: the horizon itself is not to be viewed as being supported by this type of hair. See \cite{Dias:2018yey} for a related discussion. 

Our main result in this paper is to show that there exist non-trivial solutions of the Einstein-Maxwell-scalar system in 2+1 dimensional flat space in a box, and show that the phase space has an intricate structure. Removing either the box boundary or the scalar removes all these solutions. 

\section{The Setup}

The action for the Einstein-Maxwell-Scalar theory in 2+1-d is given by
\bea
S= \dfrac{1}{16\pi G}\int d^{3}x \sqrt{-g} \Bigg(R - F_{\mu\nu}F^{\mu\nu} - |\nabla_{\mu} \psi-i q A_{\mu} \psi|^{2} \Bigg) + \dfrac{1}{8\pi G}\int d^{2}x \sqrt{-\gamma} K,
\eea
where $ \gamma $ is the boundary metric and $ K $ is the extrinsic curvature.

We work with the metric ansatz \cite{Hartnoll:2008kx, BKA}
\bea
ds^{2} = -g(r) h(r) dt^{2} +\dfrac{dr^{2}}{g(r)} +r^{2}d\theta^{2},
\eea
with $\theta \sim \theta +2 \pi$, and for the Maxwell field, we choose the gauge
\bea
A_{\mu} = (\phi(r),0,0).
\eea
In this gauge, we find that the complex scalar can be chosen to be $ \psi = \psi(r)e^{i\alpha} $, where $ \alpha $ is some constant and doesn't play a role in the analysis and $ \psi(r) $ is real.  With this ansatz, we get the equations of motion
\bea
\frac{g'(r)}{2 r g(r)}+\frac{q^2 \psi (r)^2 \phi (r)^2}{2 g(r)^2 h(r)}+\frac{\phi '(r)^2}{g(r) h(r)}+\frac{1}{2} \psi '(r)^2=0,\\
\frac{h'(r)}{2}-r h(r) \psi '(r)^{2}-\frac{q^2 r \psi (r)^2 \phi (r)^2}{g(r)^2}= 0,\\
\phi ''(r)+\frac{\phi '(r)}{r}-\frac{q^2 \psi (r)^2 \phi (r)}{2 g(r)}-\frac{h'(r) \phi '(r)}{2 h(r)}=0,\\
\psi ''(r)+\frac{\psi '(r)}{r}+\frac{g'(r) \psi '(r)}{g(r)}+\frac{q^2 \psi (r) \phi (r)^2}{g(r)^2 h(r)}+\frac{h'(r) \psi '(r)}{2 h(r)}=0,
\eea
where the first two are from the variation of the metric, third from variation of the gauge potential and the last one from the variation of the scalar.

We will specify the theory by specifying the boundary metric. In 2+1 dimensions, we will take it to be
\bea
ds^2=-dt^2+r_b^2 d\theta^2 \label{box}
\eea
This is our box. Note in particular that $r=r_b$ is the boundary. Loosely speaking, specifying a boundary is necessary because gravity is a holographic theory, see \cite{PCB} for some parallel explanations in 3+1 dimensions with some technical detail. 

The scaling symmetries of the equations of motion are
\begin{itemize}
	\item $ r\rightarrow a r $, $ q\rightarrow q/a $, $ t\rightarrow at $ : with this the line element scales as $ ds^{2}\rightarrow a^{2}ds^{2} $, the scale is set such that $ \rb =1 $.
	\item $ h\rightarrow a^{2}h $, $ \phi\rightarrow a \phi $, $ t\rightarrow t/a $: this is used to set $ g(\rb)h(\rb)=1 $.
	\item $ q\rightarrow \sqrt{a}q $, $ g\rightarrow a g $, $ h\rightarrow h/a $: this we can use to set $ g(r=0)=1 $.
\end{itemize}
Versions of the first two scaling symmetries were present in both AdS as well as in the higher dimensional box, but the last scaling symmetry is unique to 2+1 dimensions. 
	
\section{Boson Star}

We find that the boson star is obtained as a second order phase transition from the empty box background. For these solutions, we define the boundary conditions at $ r=0 $ to be
\bea
\psi'(0) =0,~~~ \phi'(0) =0,~~~ h'(0) = 0, ~~~g'(0)= 0. 
\eea
These are just regularity conditions for the fields at the origin.

With this, we can do a Taylor series expansion around $ r=0 $ to get
\bea
g(r) = g_0 - \dfrac{\phi_{0}^2 \psi_{0}^2 q^2 r^2}{2 g_0 h_{0}}+\dots,\\
h(r) = h_{0} + \dfrac{\phi_{0}^2 \psi_{0}^2 q^2 r^2}{g_0^{2}} +\dots,\\ 
\phi(r) = \phi_{0} + \dfrac{1}{8g_0} \phi_{0} \psi_{0}^2 q^2 r^2+ \dots,\\
\psi (r) = \psi_{0} - \dfrac{\phi_{0}^2 \psi_{0} q^2 r^2}{4 g_0^{2} h_{0}} + \dots,
\eea
where $ \psi(0) = \psi_{0} $, $ \phi(0) =\phi_{0} $, $g(0) = g_0 $ and $ h(0) = h_{0} $. 

One thing worth noting here is that unlike in higher dimensions, we have a {\em choice} in 2+1 dimensions to pick a value for $g_0$. In higher dimensions, the choice $g_0=1$ is forced on us by the equations of motion both in flat space as well as in AdS. This (among other things) means that the phase diagram of boson stars that we will see here has more parameters than in higher dimensions. The technical reason for this difference at the level of the equations of motion is that the angular direction of the geometry in the present case is a circle and therefore carries no curvature. However, with the scaling symmetry with which we can set $ g(r=0) =1 $, we can indeed study the boson star configurations for different values of $ q $. Also what the scaling symmetry allows is to map a configuration with arbitrary $ g(0) $ and $ q $ to $ g(0)=1 $ and an appropriate value of $ q $ obtained by using the scaling.

We will not aim for an exhaustive study of the phase diagram, but we will explore a few different choices of $g_0$, mainly $g_0=1$. This is interesting because it helps us compare easily with higher dimensional examples: it is also interesting to study these phase diagrams where all quantities are fixed at the boundary, but we will not undertake this in this paper. But we emphasize that by using the re-scaling symmetries on our solutions, we can reach the full solution space. We will further restrict our attention to the case where the scalar field at the boundary is fixed (Dirichlet) to be zero: this is the case which can be naturally compared to a boson star with a non-trivial profile in the interior. This choice was also made in \cite{PCB}.\footnote{It will be interesting to also construct solutions that satisfy Neumann boundary conditions for the metric at the boundary, along the lines of \cite{Neumann1,Neumann2,Neumann3}.}

\subsection{Boson Star Instability}

To find the parameters where the transition to boson star occurs, we analyse the scalar equation of motion as in a background where $ g(r) =g_0 $, $ h(r) =h_0 $ and $ \phi(r) =\phi_0 $. The scalar eom in this probe limit is given by 
\bea
\psi''(r) + \dfrac{\psi'(r)}{r} + \dfrac{q^{2}\phi_0^{2}}{g_0^{2}h_0} \psi(r) =0.
\eea
The solutions with the boundary condition $\psi'(0) = 0 $ and $ \psi(\rb =1 ) = 0 $ is obtained when we have
\bea
\psi(r) = \psi_0 ~J_0 \left(\dfrac{q ~\phi_0}{g_0 \sqrt{h_0}} r\right),
\eea
where $\psi_0 = \psi(0)$ and $J_0$ is the Bessel Function of first kind with index 0. The function has zeroes at $\rb =1$, when the argument is such that
\bea
\dfrac{q ~\phi_0}{g_0 \sqrt{h_0}} \simeq 2.4048.., 5.520, \dots
\eea
where $ \phi_0 $ is the critical value of gauge field where the boson star instability occurs, and is related to the critical charge density\footnote{Note that unlike in higher dimensions, the quantity we are holding fixed is a charge density and not a chemical potential, see \cite{Princeton}. The reason is that the gauge field in 2+1 dimensions goes asymptotically as $C_1 \log r + C_2$: the leading behavior which is more naturally treated as the chemical potential goes as $\log r$.} for the instability $\rho_c = \frac{\phi_{0}}{\sqrt{g_{0}h_{0}}} $.

\begin{figure}[H]
\centering
	\includegraphics[scale=.7]{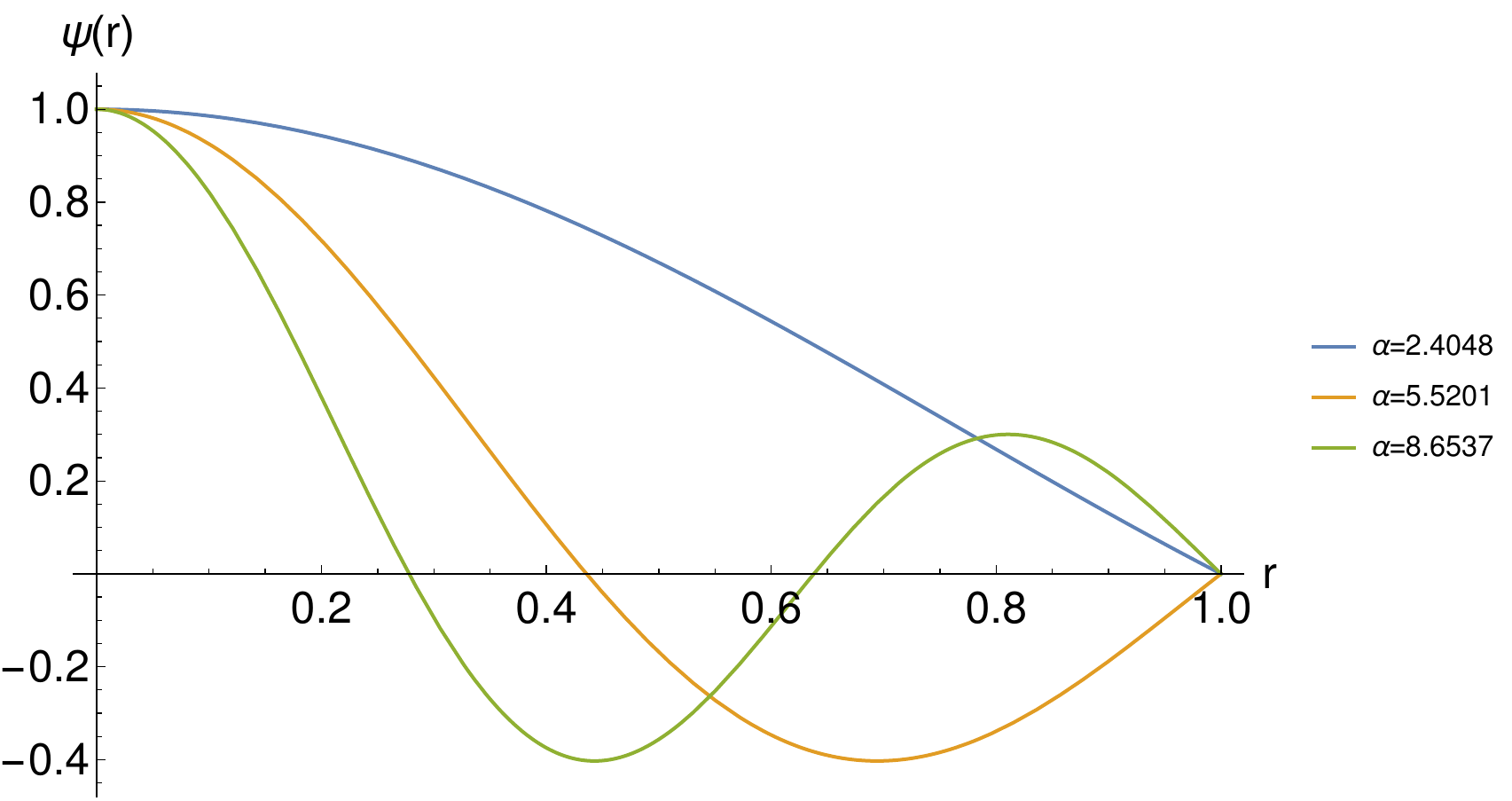}
	\caption{$ r $ vs $ \psi(r) $ plot for boson star instability for different $\alpha = \frac{q \phi_0}{g_0 \sqrt{h_0}}$.}
	\label{profiles}
\end{figure}

\begin{figure}[H]
\centering
	\includegraphics[scale=.7]{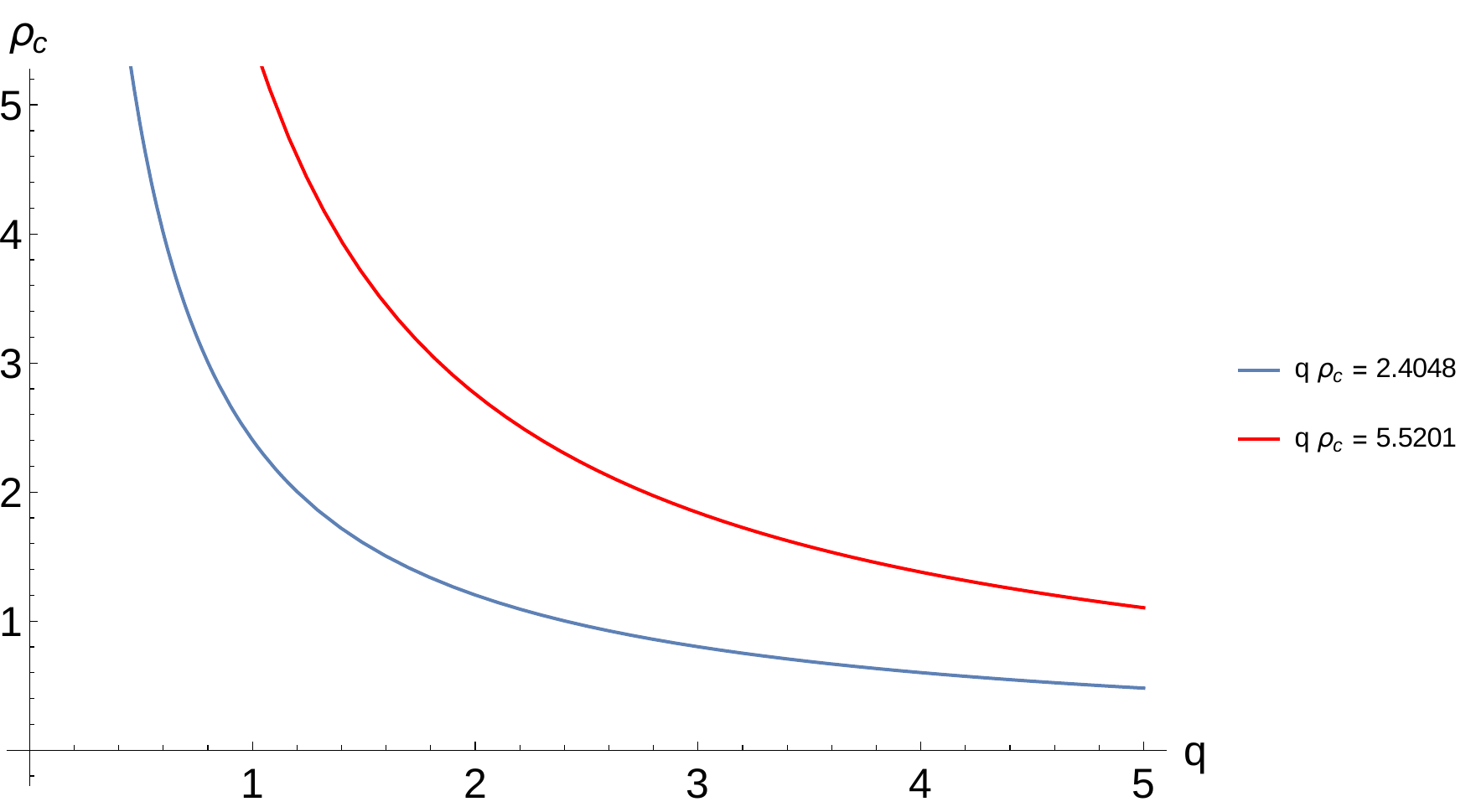}
	\caption{$ q $ vs $ \rho_{c} $ plot for boson star instability, with $g_0 =h_0 =1$.}
	\label{instabilityplot}
\end{figure}

The profiles for first 3 distinct solutions is given in Fig.\ref{profiles}. The plot $ q $ vs $ \rho_{c} $ for the boson star instability is given in Fig.\ref{instabilityplot}. The thermodynamic stability of the solutions will be discussed when we look at the free energy.

\subsection{Backreacted Solutions}

The fully backreacted solutions are found by giving values to the free parameters $\psi_0,~ \phi_0, ~ g_0 $ and $h_0$, for some given $q$, and integrating out to the boundary. Now $ h_{0} $ can be taken to be arbitrary, say $ h_{0}=1 $, as we will rescale $ h(r) $ at the end using the scaling symmetry mentioned above. We also can set $ g_{0} =1 $ using the third scaling symmetry. So for a given value of $ q $, the free parameters at $ r=0 $ are $ \psi_{0} $ and $ \phi_{0} $. The set of equations of motion are solved by numerical shooting. Here, for a given value of $q$, for different values of $ \psi_{0} $, we tune $ \phi_{0} $ such that $ \psi(\rb) =0 $. This process is then repeated for different choices of $q$. The charge density is obtained as $ \rho = \phi(\rb) $, after rescaling $ \phi $ and $ h $ such that $ g(\rb)h(\rb) =1 $. 

The two interesting quantities at the boundary are $ \rho $ and $ |\psi'(\rb)| $, a plot of which for different values of $ q $ and $g_0$\footnote{We have looked at different $ g_0 $ as well, so that we can have a clearer picture of the system.} are given in Fig.\ref{condensateplots}. As we increase the value of $\psi_0$, the value of $|\psi'(\rb)|$ also increases (while holding $\psi(\rb)=0$), so increasing the value of $\psi_0$ can be understood as going up the corresponding curve in the $ \rho $ vs $ |\psi'(\rb)| $ plots.

\begin{figure}[H]
    \centering
    \begin{subfigure}{0.7\textwidth}
        \includegraphics[width=\textwidth]{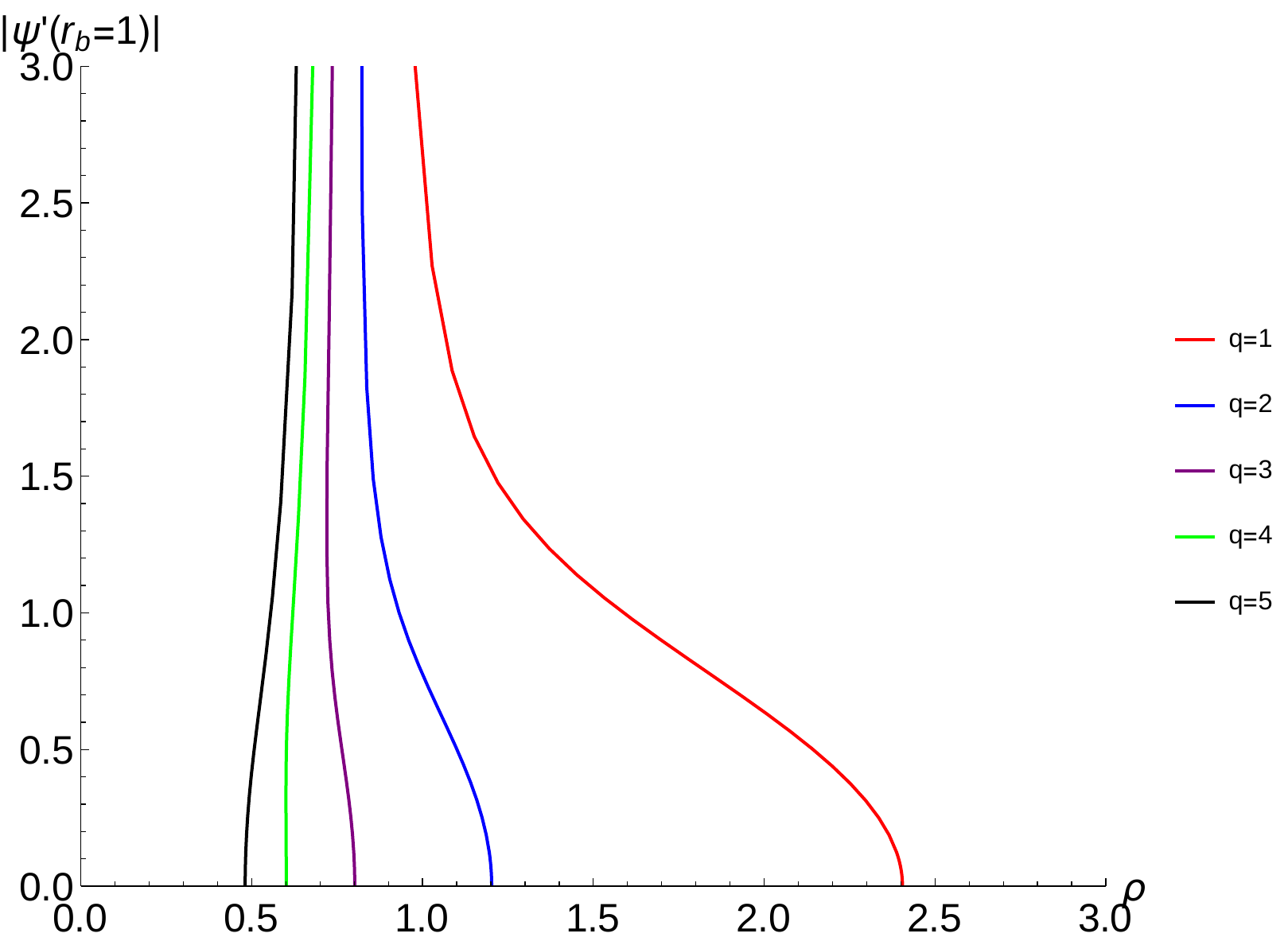}
        \caption{$g_0 = 1$}
        \label{psiprofilebs}
    \end{subfigure}\end{figure}
    
    \begin{figure}[H]\ContinuedFloat
    \centering
    \begin{subfigure}{0.7\textwidth}
        \includegraphics[width=\textwidth]{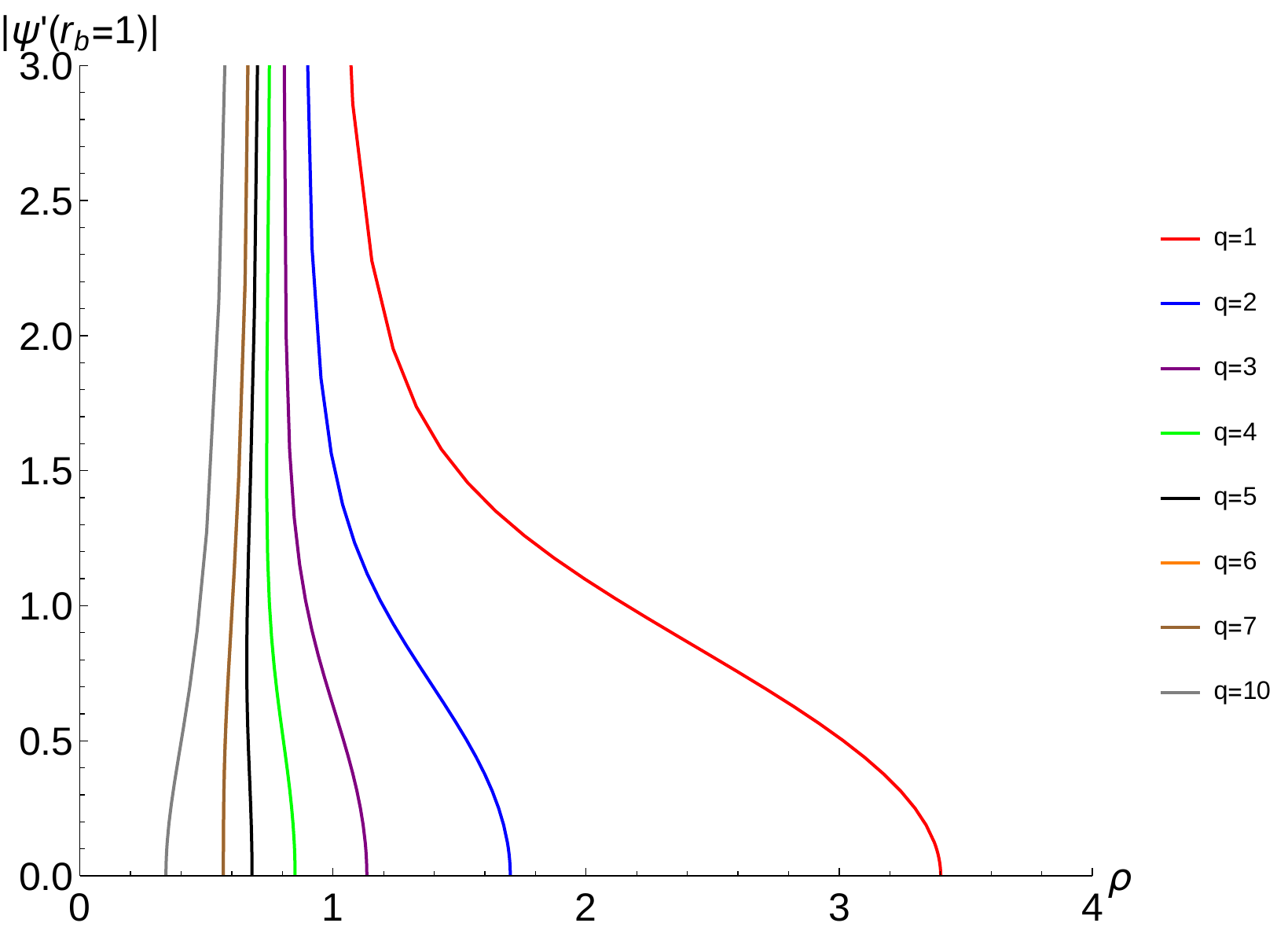}
        \caption{$g_0$ = 2}
        \label{phiprofilebs}
    \end{subfigure}
    \end{figure}
    
    \begin{figure}[H]\ContinuedFloat
    \centering
    \begin{subfigure}{0.7\textwidth}
        \includegraphics[width=\textwidth]{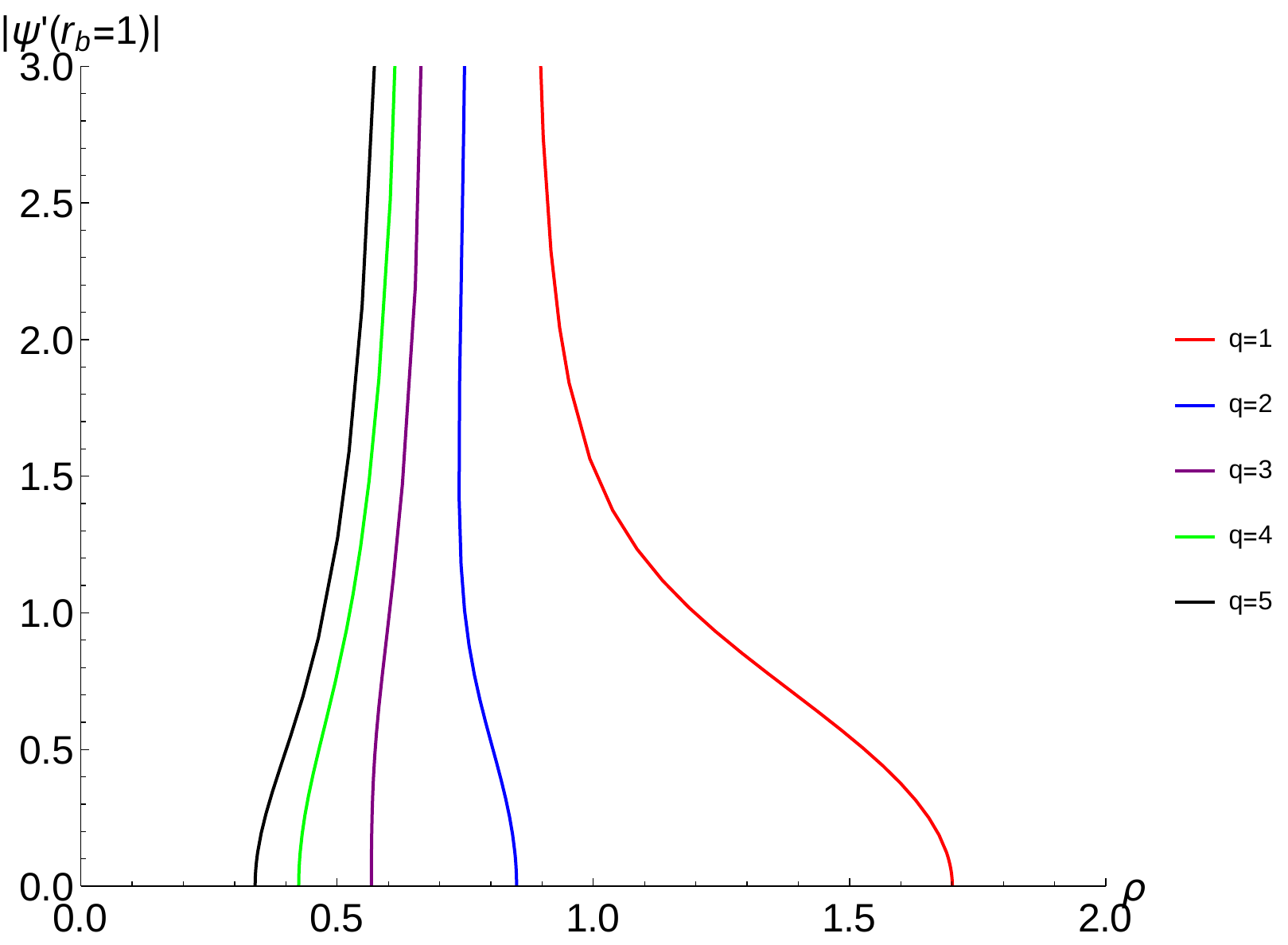}
        \caption{$g_0 = \frac{1}{2}$ }
        \label{phiprofilebs}
    \end{subfigure}
     \caption{$\rho$ vs. $\psi'(\rb=1)$ for different values of $g_0$ and $q$}\label{condensateplots}
\end{figure}

\begin{figure}[H]
\centering
	\includegraphics[scale=.7]{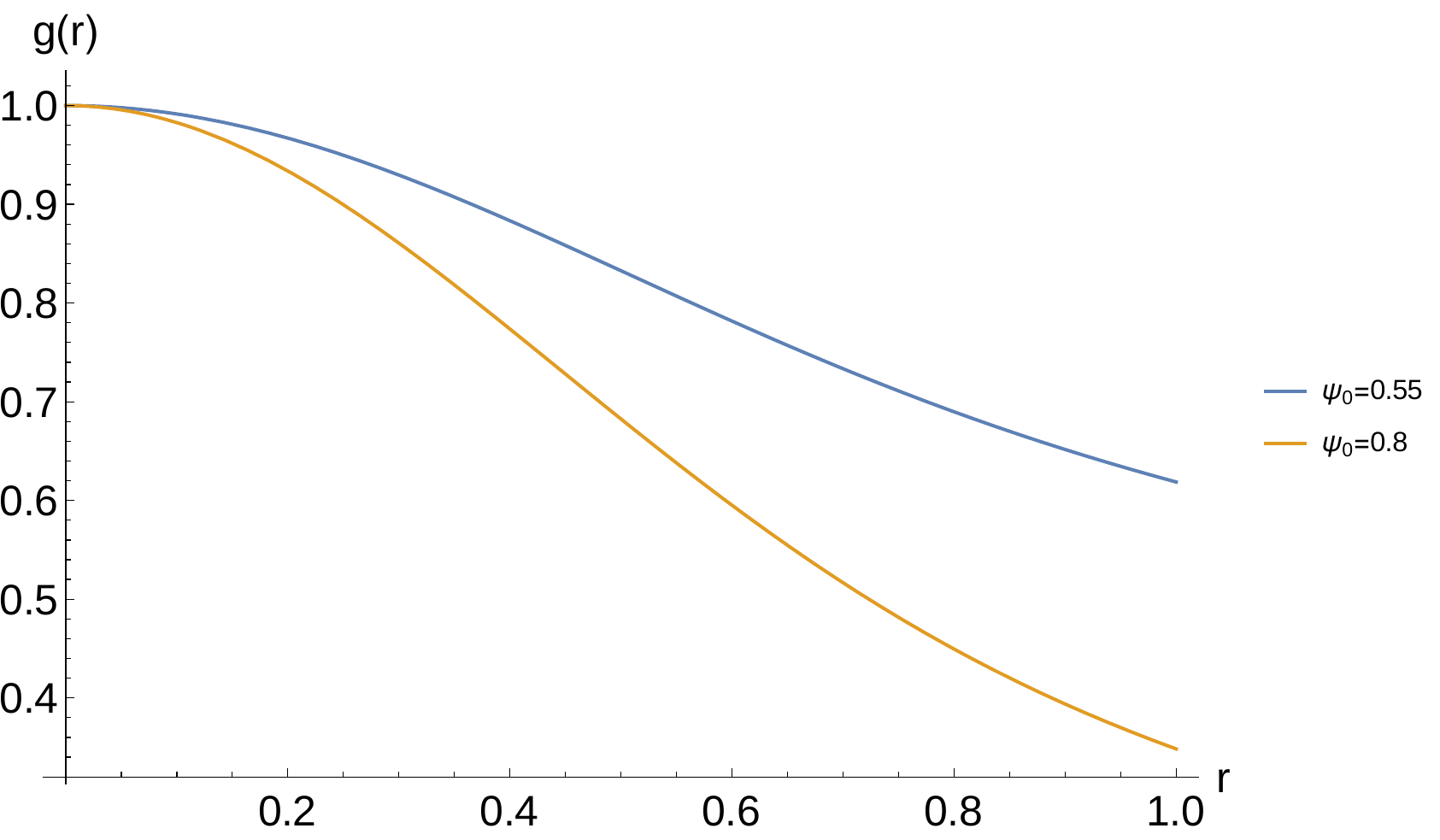}
	\caption{$ r $ vs $ g(r) $ plots for $ q=1 ,~ g_0=1$ and $ \psi_{0}=0.55,0.8 $.}
	\label{gplots}
\end{figure}

The point to note is: suppose we work with, say $g_0 =1 $ and $ q=1 $, then the profile of $ g(r) $ for $ \psi_{0} = 0.55,0.8 $ is given in Fig.\ref{gplots}. If we keep increasing $ \psi_{0} $, then at some point $ g(\rb) = 0 $, and a further increase of $ \psi_{0} $ would result in $ g(r) =0 $ for some $ r<\rb $. Vanishing of $g$ indicates the breakdown of the box interpretation, so we will not allow it.  This means there is a maximum value of $ \psi_{0} $ that we can chose beyond which we fail to get a meaningful system.

What this means is that for a given $ q $, the charge density which supports a boson star phase is when $\rho $ is in between $ \rho_{c}$ and $\rho_{max} $, where $ \rho_{0} $ is the charge density at which the boson star instability happens for the given $ q $, and $ \rho_{max} $ is the charge density associated to the fully backreacted configuration where the $ g(\rb) $ is almost (but not) 0. The way to find $ \rho_{0} $ is very easy given that we are doing an analysis on the probe scalar equations of motion, whereas $ \rho_{max} $ is a bit challenging owing to that fact that one has to deal with the fully backreacted system and slowly tweak $ \psi_{0} $ and $ \phi_{0} $ such that we have $ g(\rb)\gtrsim 0 $, for each $ q $ separately. Also one finds that, for $ \psi_{0} $ such that $ g(\rb)\rightarrow 0 $, we will have $ |\psi'(\rb)|\rightarrow \infty $. This phenomena is visible in Fig.\ref{condensateplots}, where for each $ q $ we can see that curves will be such that $\rho $ is in between $ \rho_{c}$ and $\rho_{max} $.

\begin{figure}[H]
\centering
\includegraphics[scale=0.7]{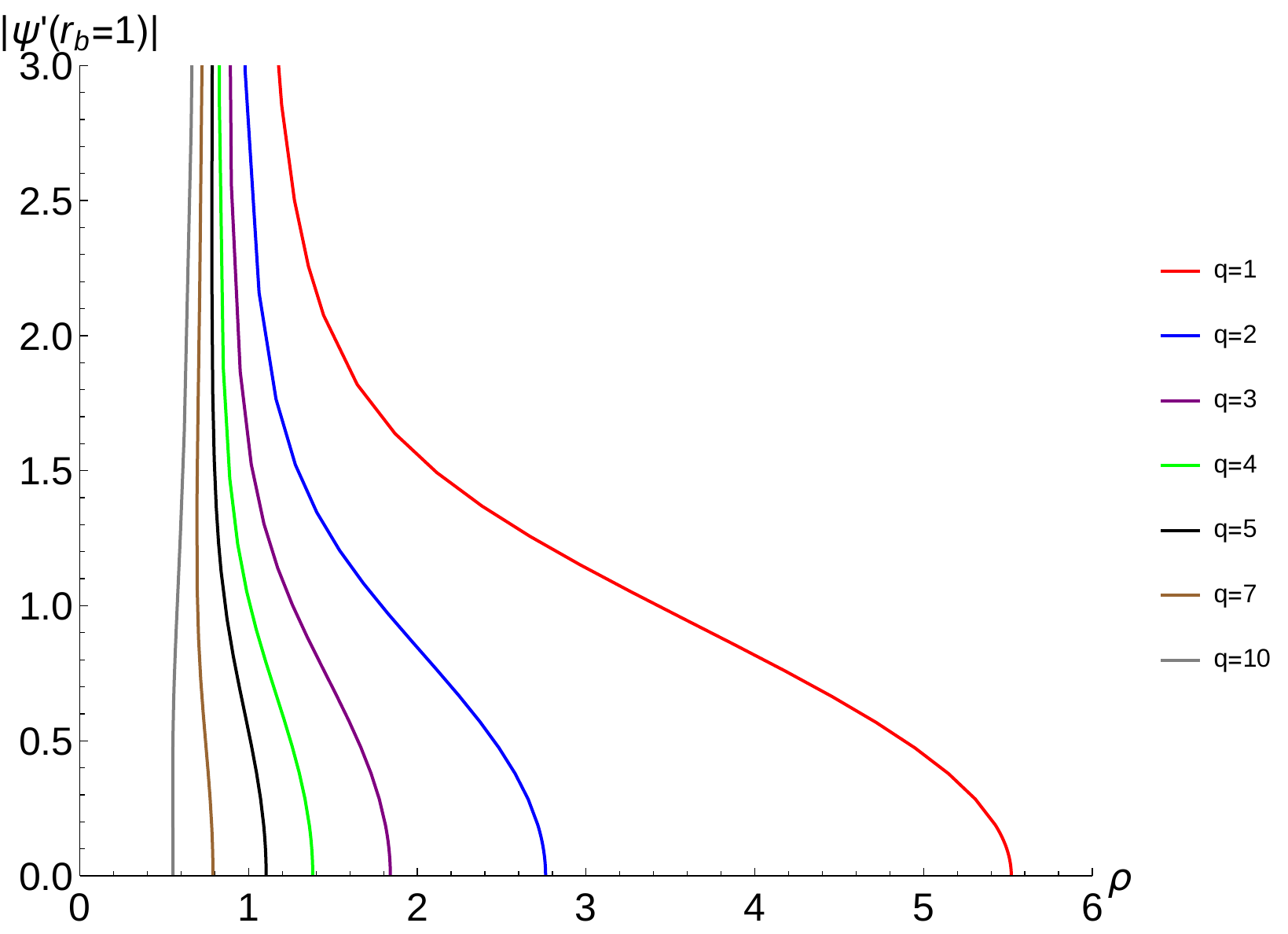}
\caption{$\rho$ vs. $|\psi'(\rb=1)|$ plot for scalar profiles with 1 node.}
\label{1nodecond}
\end{figure}

We could also work with solutions that have 1 or more nodes in the scalar profile. The $\rho$ vs $|\psi'(\rb=1)|$ plots for $g_0 = 1$ and different values of $q $ is given in Fig.\ref{1nodecond}.

\section{Phase Diagram}

\begin{figure}[H]
\centering
\includegraphics[scale=0.7]{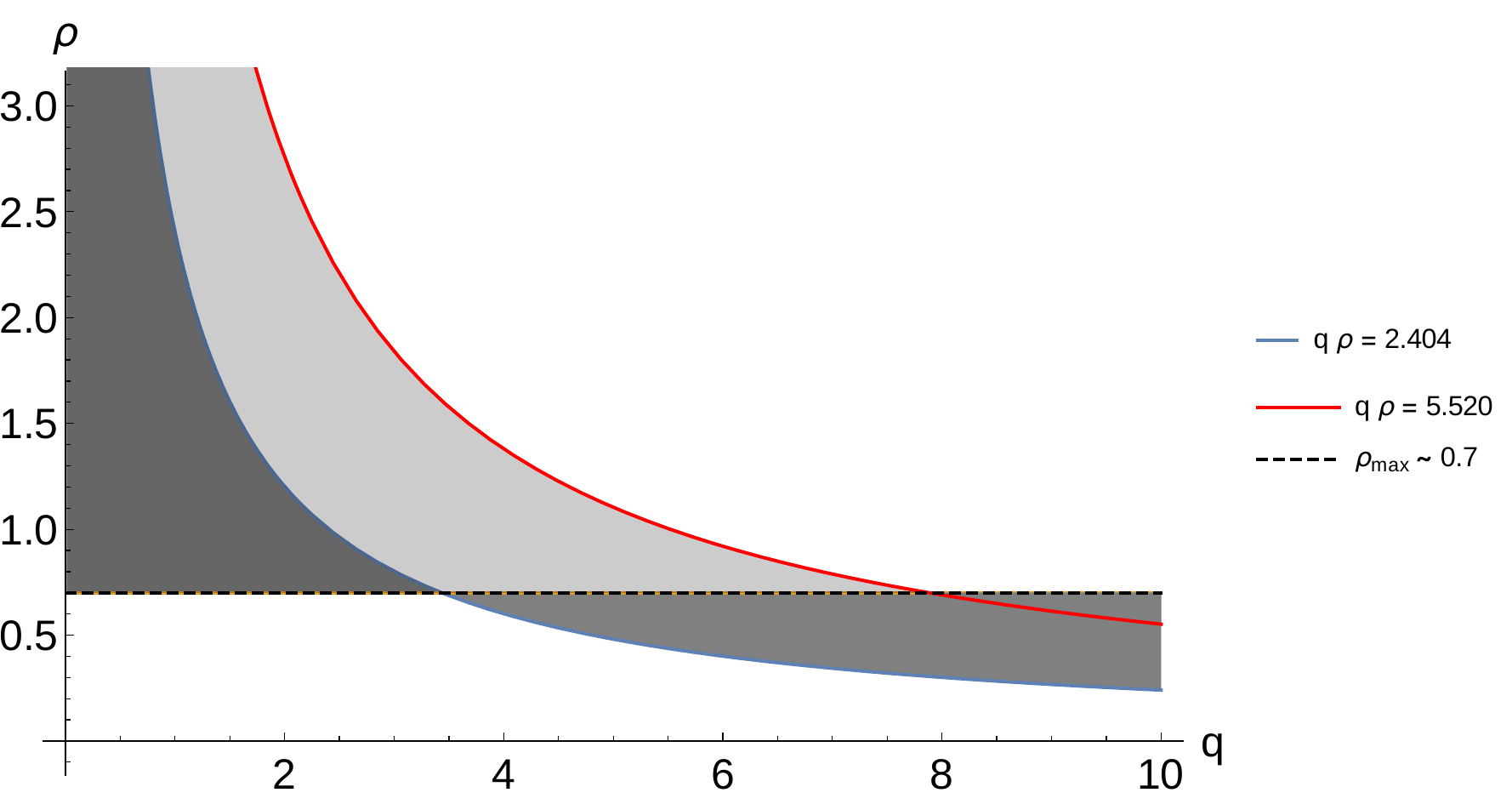}
\caption{$q$ vs. $\rho$ plot with regions shaded where boson stars with 0 and 1 nodes exist.}
\label{regions}
\end{figure}

Since the solutions we consider do not have a well defined temperature\footnote{Note that in 3+1 dimensions (as well as here) the temperature of the boson star is fixed by the periodicity of the Euclidean circle which can be chosen arbitrarily. But in 3+1 dimensions, because of the existence of the black hole, at the phase transition line one could demand that the temperature of the boson star match that of the critical hairy black hole. Here, because of the absence of black holes, there is no such possibility.}, the phase diagram we study is the region in the $q -\rho$ space where boson star solutions can exist. As we have seen in the condensate plots (Fig.  \ref{condensateplots}, \ref{1nodecond}), for a given $q$ the boson star configuration is present when $\rho$ is in between $\rho_c$ and $\rho_{max}$. Hence, we can mark-out this region in a $q-\rho$ plot as shown in Fig.\ref{regions}.
The black dotted line is an approximate curve where we expect the $\rho_{max}$ to occur. This expectation is based on comparing the curves for the 0-node and 1-node case, the $\rho_{max}$ seems to be the same (or roughly the same) for both of them, and speculatively for boson stars with any number of nodes. The region where a boson star exists is the region between the instability curve and the $\rho_{max}$.

\subsection{Free Energy}
To define the free energy, we look at the on-shell Euclidean action. The extrinsic curvature at the boundary for the metric ansatz we use is
\bea
K = \dfrac{\sqrt{g(\rb)}}{\rb} + \dfrac{g'(\rb)}{2\sqrt{g(\rb)}} + \dfrac{\sqrt{g(\rb)}~h'(\rb)}{2h(\rb)}.
\eea
We can simplify the on-shell action using the fact that $ \mathcal{L} -R  = 2 \tilde{T}_{\theta\theta}/r^{2}$, where $ \mathcal{L} $ is the Lagrangian and the Einstein equation is written in the form $ G_{\mu\nu} =\tilde{T}_{\mu\nu} $. We also work in $ G=1 $ units, and the wick rotation is defined as $ t\rightarrow -i\tau $. Trading $ \tilde{T}_{\theta\theta} $ for $ G_{\theta\theta} $, and simplifying, we get
\bea
\dfrac{1}{16\pi} \int d^{3}x \sqrt{-g} \mathcal{L}\bigg|_{on-shell} = -\dfrac{i}{16\pi} 2\pi\beta \int_{r_{0}}^{\rb} dr ~\dfrac{d}{dr}\left(-2 g(r)\sqrt{h(r)}\right) = -i\dfrac{\beta}{4}~ g(r)\sqrt{h(r)}\bigg|^{\rb}_{r_{0}}.
\eea
Rescaling $ h(r) $ to get the boundary metric of the form \eqref{box}, we get
\bea
\dfrac{1}{16\pi} \int d^{3}x \sqrt{-g} \mathcal{L}\bigg|_{on-shell} = -i \dfrac{\beta}{4}\left(\sqrt{g(\rb)} - \dfrac{g(0)\sqrt{h(0)}}{\sqrt{g(\rb)h(\rb)}}\right).
\eea
One thing to note is that the extrinsic curvature is invariant under the scaling of $ h(r) $.

The extrinsic curvature of the empty box is $ K_{0} = \frac{\sqrt{g(0)}}{\rb} $, which we subtract from the action, so that the on-shell action for the empty box is 0. With this, we get the free energy directly from the on-shell Euclidean action, 
\bea
F = \dfrac{S_{euc}}{\beta} = \dfrac{1}{4}\Bigg( \sqrt{g(0)} \Bigg(1- \sqrt{\dfrac{g(0)h(0)}{g(\rb)h(\rb)}}\Bigg) - \dfrac{\rb~ g'(\rb) }{2\sqrt{g(\rb)}} - \dfrac{\rb \sqrt{g(\rb)}~h'(\rb)}{2h(\rb)}\Bigg).
\eea
Now using the third rescaling symmetry, whereby $ g\rightarrow g_{0} g $, we can set $ g(0)=1 $ and we get
\bea
F =  \dfrac{\sqrt{g_{0}} }{4}\Bigg( 1- \sqrt{\dfrac{h(0)}{g(\rb)h(\rb)}} - \dfrac{\rb~ g'(\rb) }{2\sqrt{g(\rb)}} - \dfrac{\rb \sqrt{g(\rb)}~h'(\rb)}{2h(\rb)}\Bigg).
\eea
With this we can evaluate the free energy for arbitrary $ q $ and $ g_{0} $, if we know the free energy in $ g(0) =1 $ ensemble for the corresponding value of $ q $. 
The factor inside the parenthesis is the free energy of the $ g(0)=1 $ ensemble.

\begin{figure}[H]
	\centering
	\includegraphics[scale=0.7]{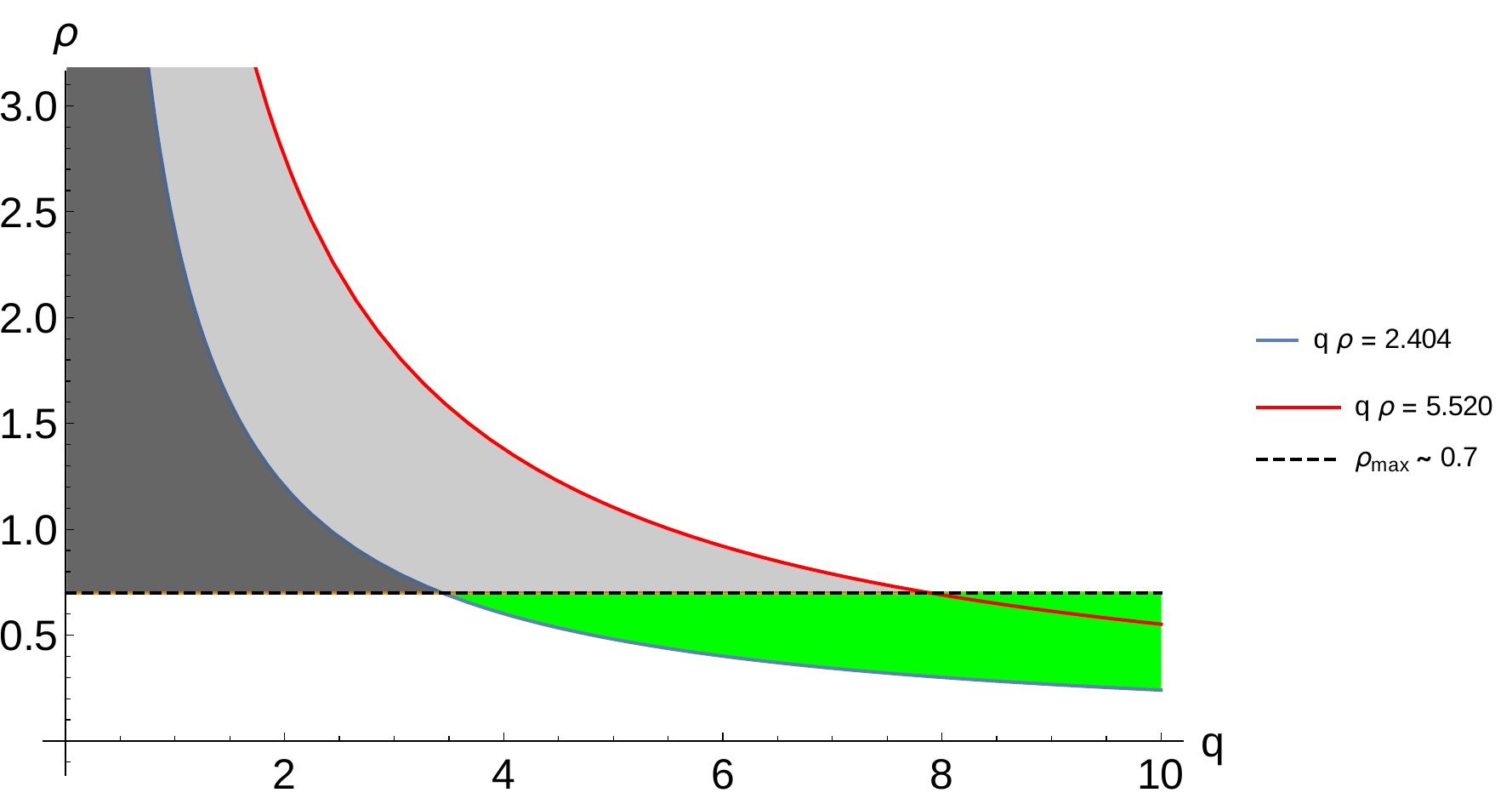}
	\caption{$q$ vs. $\rho$ plot with $ g_{0} =1 $, green region depicting the thermodynamically stable 0-node configuration.}
	\label{regions1}
\end{figure}

Now, we can evaluate the free energy of different boson star configurations. When we look at the $ g(0) $ fixed ensemble (we will focus only on $ g(0)=1$ for the reasons mentioned earlier), the free energy of the boson star configurations are negative, meaning thermodynamically stable, only when the charge of the scalar is such that $ \rho_{max}> \rho_{c} $. In other words, this happens only for $ q $ greater than the value at which the instability curve and the $ \rho_{max} $ line cross-over\footnote{A similar phenomenon was also noted in 3+1 dimensions in \cite{PCB}.}. This is checked to be true for the 0 and 1 node solutions. However, in the region where the boson star configurations are present, there always exist a 0-node boson star solution, which will be thermodynamically more stable. Hence, we can say that the only thermodynamically stable configurations are the 0-node solutions. In Fig.\ref{regions1}, the green region marks out the parameter space where the 0-node boson stars that are thermodynamically stable exist.

%

\section{Black Holes}

We will first present the explicit calculation in a spherically symmetric setting to get some intuition about why a black hole might be forbidden in flat space even inside a box. Then we will present the more general argument which is an adaptation of Hawking's horizon topology theorem to 2+1 dimensions.

\subsection{Radially Symmetric Cases}

To get some intuition for understanding the (lack of) existence of hairy black holes in the 2+1-d box, it is useful to work with a more general action that allows a symmetry-breaking potential for the scalar, as well as a cosmological constant. The basic reason for this is that the minimum potential energy of the scalar acts as a tunable contribution to the cosmological constant, enabling us to see the emergence of various features quite explicitly as we move from negative to positive $\Lambda_{eff}$. So we consider
\bea
S= \dfrac{1}{16\pi G}\int d^{3}x \sqrt{-g} \Bigg(R- 2\Lambda - F_{\mu\nu}F^{\mu\nu} - |\nabla_{\mu} \psi-i q A_{\mu} \psi|^{2} + \dfrac{m^{2}}{2}\psi^{2} - \dfrac{\lambda}{24} \psi^{4}\Bigg).
\eea
The boundary conditions at the horizon $ \rh $ are given by $ g(\rh) = 0 $, $ \phi(\rh) = 0 $. By solving the equations of motion order by order
\bea
g(r) &=& (r-\rh) \left(-\frac{2 \phi_{0}^2 \rh}{h_{0}}-2\Lambda \rh+ \dfrac{m^2}{2} \psi_{0}^2 \rh-\frac{1}{24} \lambda  \psi_{0}^4 \rh\right) + \dots,\label{grhexpand}\\
h(r) &=& h_{0} + \frac{8 h_0^2 \psi _0^2  \left(h_0 \left(\lambda  \psi _0^2-6 m^2\right){}^2+144 q^2 \phi _0^2\right)}{r_h \left(h_0 \left(\lambda  \psi _0^4+48 \Lambda -12 m^2 \psi _0^2\right)+48 \phi _0^2\right){}^2} \left(r-r_h\right)+ \dots,\\
\phi(r) &=& \phi_{0} (r-\rh) +\dots,\\
\psi(r) &=& \psi_{0} -\frac{2 h_0 \psi _0  \left(\lambda  \psi _0^2-6 m^2\right)}{r_h \left(h_0 \left(\lambda  \psi _0^4+48 \Lambda -12 m^2 \psi _0^2\right)+48 \phi _0^2\right)} (r-\rh)+\dots ,
\eea
where $ \psi(\rh) = \psi_{0} $, $ \phi'(\rh) = \phi_{0} $ and $ h(\rh) = h_{0} $. 

Now, let us look at the potential term more closely. The effective potential in the theory is 
\bea
V_{eff}(\psi) = -\dfrac{m^{2}}{2}\psi^{2} + \dfrac{\lambda}{24} \psi^{4} + 2\Lambda.
\eea
The minima of this potential is at $ \psi_{min} = \pm \sqrt{\frac{6m^{2}}{\lambda}} $, and the value of the potential is 
\bea
V_{eff}(\psi_{min}) = -\dfrac{3m^{2}}{2\lambda} + 2\Lambda.
\eea
Now we choose $ \Lambda $, such that we have $ V_{eff}(\psi_{min}) = 0 $\footnote{Note that if we allow negative values for the minimum of $V_{eff}$ at the horizon, we can introduce a local negative cosmological ``constant" and violate the dominant energy condition. We wish to avoid this scenario in the present paper. Let us note however that if one allows negative potentials, a rich space of domain wall like black hole solutions can be found that interpolate between a negative potential value at the horizon and the maximum at the box \cite{future}. This scenario is certainly interesting because $\Lambda=0$ belongs to this category. But we will not consider this case in the present paper, because if we take the box to infinity, the ground state is now AdS.}, which gives
\bea
\Lambda = \dfrac{3m^{2}}{4\lambda}.
\eea
In the expansion of $ g(r) $ around $ \rh $ given in \eqref{grhexpand}, the coefficient of $ (r-\rh) $ is
\bea
\left(-\frac{2 \phi_{0}^2 \rh}{h_{0}}-2\Lambda \rh+ \dfrac{m^2}{2} \psi_{0}^2 \rh-\frac{1}{24} \lambda  \psi_{0}^4 \rh\right) = \left(-\frac{2 \phi_{0}^2 \rh}{h_{0}} -\rh V_{eff}(\psi_{0})\right).
\eea
Since we have imposed the condition that $ V_{eff}(\psi)\geqslant 0 $ by the specific choice of $ \Lambda $, the second term is always non-positive. The first term is also non-positive, becoming 0 when $ \phi_{0} = 0 $. This means that the coefficient of $ (r-\rh) $ is always non-positive, which is not desirable if we are to get a hairy black hole solution, as $ g'(\rh) \leqslant 0  $. We see very explicitly that when the effective cosmological constant becomes negative, a black hole can exist.

One might think that a possible way out of this, is to consider the possibility of an extremal black hole, where not just $g(r_h)=0$, but also $g'(r_h)=0$. It is clear that if we try to set $ \phi_{0} = 0 $ and $ \psi_{0} = \psi_{min} $ in the above ``near-horizon'' series expansion, so that we get $ g'(\rh) = 0 $, the denominators of coefficients of $ (r-\rh) $ of the functions $ h(r) $ and $ \psi(r) $ will go to zero and the series fails to make any sense. But is it possible that by trying a power-series expansion where $ g'(\rh) = 0 $ from the outset, we could get a new form for the series expansion that bypasses this problem? We have checked this possibility as well, and it also does not lead to a viable perturbative solution. 



\subsection{The General No-Go Argument}

The discussion above had a very ``rigid'' feel, even though it was restricted to the case of spherical symmetry, so one might wonder if there exists a theorem that rules out horizons altogether in 2+1 dimensions. In fact, it is clear from the explicit calculations that it is the (absence of) curvature contribution from the compact circle dimension that results in the difficulty in obtaining a horizon\footnote{In AdS, the AdS length scale acts as a proxy for this, and therefore the BTZ black hole manages to exist \cite{BTZ}.}. So one might suspect that an argument that ties together horizon topology and curvature via a positivity condition on energy might be at play here.

A natural place to look for such an argument is Hawking's horizon topology theorem. In its original incarnation \cite{Hawking} it was used to show that under the dominant energy condition and a few other natural assumptions, horizon topologies in 3+1 dimensions must be 2-spheres. In \cite{Oz} it was generalized to higher dimensions. A crucial and (promising, from our perspective) ingredient in these theorems is that they use dominant energy condition, and therefore naturally do not apply in asymptotically AdS geometries. 

The basic strategy of these theorems is to show that under the assumption of dominant energy condition, the existence of a horizon one wishes to rule out, will imply the existence of a (marginally outer) trapped surface slightly outside the horizon, which is forbidden\footnote{The idea here is that the horizon contains all that is not visible from asymptotic infinity, and therefore a trapped surface must also be within the horizon, and therefore cannot be outside it. This argument is clean when we work with asymptotically flat spaces, but since we have defined the box in a specific coordinate system, it is less clean in our case. However, since the argument leads to a trapped surface arbitrarily close to the horizon (but outside it), we expect that this cannot substantively change the conclusion that there are no black holes.}. The key condition for our purposes is eq. (2.1) in \cite{Oz} 
\bea
\int_{M_H} \hat {\cal R} d\hat S > 0
\eea
where the integral is over the event horizon, and the Ricci curvature is with respect to the induced metric on the horizon. Note that this condition is independent of dimension. In 3+1 dimensions, it leads to the restriction that the genus of the horizon must be zero, implying spherical horizons \cite{Hawking}. In higher dimensions, it leads to related but distinct constraints \cite{Oz}.  In 2+1 dimensions however, it is clear that this condition can never be satisfied for both compact or non-compact horizons, because the only 1-dimensional manifolds are lines and circles neither of which have curvature under any metric. This immediately rules out the possibility of black holes in 2+1 dimensions\footnote{Let us emphasize again however, the caveats noted in the previous subsection.}. A proof of this statement is known in the literature \cite{NoGo}, but here we have emphasized the point of view that it is a horizon topology argument \cite{Hawking}.

Note that the above argument is an effectively local argument around the horizon, if we do not allow trapped surfaces outside horizons. If it were violated, it would allow the possibility of trapped surfaces close to, but outside the horizon. This is consistent with our explicit calculations in the last section, where we came to similar conclusions via a near-horizon expansion without invoking any explicit statement about far-away regions, eg. the boundary. The AdS case evades the No-Go argument because it violates the dominant energy condition.



\begin{thebibliography}{99}

  \bibitem{HP} 
 S. W. Hawking and D. N. Page,
  {\em Thermodynamics Of Black Holes In Anti-De Sitter Space},
Commun. Math. Phys. {\bf 87}, 577 (1983).

\bibitem{CK}
C.~Krishnan,
``Bulk Locality and Asymptotic Causal Diamonds,''
arXiv:1902.06709 [hep-th]

\bibitem{PCB}
P.~Basu, C.~Krishnan and P.~N.~B.~Subramanian,
``Hairy Black Holes in a Box,''
JHEP {\bf 1611}, 041 (2016),
doi:10.1007/JHEP11(2016)041,
[arXiv:1609.01208 [hep-th]].


\bibitem{PCB-ads}
P.~Basu, C.~Krishnan and P.~N.~Bala Subramanian,
 ``Phases of Global AdS Black Holes,''
JHEP {\bf 1606}, 139 (2016),
doi:10.1007/JHEP06(2016)139,
[arXiv:1602.07211 [hep-th]].

\bibitem{BTZ} 
M.~Banados, C.~Teitelboim and J.~Zanelli,
``The Black hole in three-dimensional space-time,''
Phys.\ Rev.\ Lett.\  {\bf 69}, 1849 (1992)
doi:10.1103/PhysRevLett.69.1849
[hep-th/9204099].

\bibitem{chargedBTZ} 
G.~Clement,
``Spinning charged BTZ black holes and selfdual particle - like solutions,''
Phys.\ Lett.\ B {\bf 367}, 70 (1996),
doi:10.1016/0370-2693(95)01464-0,
[gr-qc/9510025].


\bibitem{Winstanley1} 
S.~R.~Dolan, S.~Ponglertsakul and E.~Winstanley,
{\em Stability of black holes in Einstein-charged scalar field theory in a cavity},
Phys.\ Rev.\ D {\bf 92}, no. 12, 124047 (2015)
[arXiv:1507.02156 [gr-qc]].

\bibitem{Winstanley2} 
S.~Ponglertsakul, S.~Dolan and E.~Winstanley,
{\em Black hole solutions in Einstein-charged scalar field theory},
arXiv:1507.02462 [gr-qc].

\bibitem{Peng1} 
Y.~Peng,
``Studies of a general flat space/boson star transition model in a box through a language similar to holographic superconductors,''
JHEP {\bf 1707}, 042 (2017),
doi:10.1007/JHEP07(2017)042,
[arXiv:1705.08694 [hep-th]].

\bibitem{Hawking} 
S.~W.~Hawking and G.~F.~R.~Ellis,
``The Large Scale Structure of Space-Time,''
doi:10.1017/CBO9780511524646.\\
S.~W.~Hawking,
``Black holes in general relativity,''
Commun.\ Math.\ Phys.\  {\bf 25}, 152 (1972),
doi:10.1007/BF01877517.

\bibitem{Oz} 
C.~Helfgott, Y.~Oz and Y.~Yanay,
``On the topology of black hole event horizons in higher dimensions,''
JHEP {\bf 0602}, 025 (2006),
doi:10.1088/1126-6708/2006/02/025,
[hep-th/0509013].

\bibitem{NoGo} 
D.~Ida,
``No black hole theorem in three-dimensional gravity,''
Phys.\ Rev.\ Lett.\  {\bf 85}, 3758 (2000),
doi:10.1103/PhysRevLett.85.3758,
[gr-qc/0005129].

\bibitem{Dias:2018yey} 
O.~J.~C.~Dias and R.~Masachs,
``Evading no-hair theorems: hairy black holes in a Minkowski box,''
Phys.\ Rev.\ D {\bf 97}, no. 12, 124030 (2018),
doi:10.1103/PhysRevD.97.124030,
[arXiv:1802.01603 [gr-qc]].

\bibitem{Hartnoll:2008kx} 
S.~A.~Hartnoll, C.~P.~Herzog and G.~T.~Horowitz,
``Holographic Superconductors,''
JHEP {\bf 0812}, 015 (2008),
doi:10.1088/1126-6708/2008/12/015,
[arXiv:0810.1563 [hep-th]].  

\bibitem{BKA} 
D.~Arean, P.~Basu and C.~Krishnan,
{\em The Many Phases of Holographic Superfluids},
JHEP {\bf 1010}, 006 (2010)
doi:10.1007/JHEP10(2010)006
[arXiv:1006.5165 [hep-th]].


\bibitem{Neumann1}
C.~Krishnan and A.~Raju, ``A Neumann Boundary Term for Gravity,''
Mod.\ Phys.\ Lett.\ A {\bf 32}, no. 14, 1750077 (2017),
doi:10.1142/S0217732317500778,
[arXiv:1605.01603 [hep-th]].

\bibitem{Neumann2}
C.~Krishnan, K.~V.~P.~Kumar and A.~Raju,
``An alternative path integral for quantum gravity,''
JHEP {\bf 1610}, 043 (2016),
doi:10.1007/JHEP10(2016)043,
[arXiv:1609.04719 [hep-th]].

\bibitem{Neumann3} 
C.~Krishnan, A.~Raju and P.~N.~B.~Subramanian,
``Dynamical boundary for anti–de Sitter space,''
Phys.\ Rev.\ D {\bf 94}, no. 12, 126011 (2016),
doi:10.1103/PhysRevD.94.126011,
[arXiv:1609.06300 [hep-th]].

\bibitem{Princeton} 
J.~Ren,
``One-dimensional holographic superconductor from AdS$_3$/CFT$_2$ correspondence,''
JHEP {\bf 1011}, 055 (2010),
doi:10.1007/JHEP11(2010)055,
[arXiv:1008.3904 [hep-th]].


\bibitem{future} ``A  Black Hole in Three Dimensions without a Cosmological Constant", to appear.

\end{thebibliography}
\end{document}